\documentclass[12pt]{article}

\usepackage{amsmath}
\usepackage{amsfonts}
\usepackage{eufrak}
\usepackage{amssymb,bbold}
\usepackage{epsfig}
\usepackage{bm}

\def\be{\begin{equation}}
\def\ee{\end{equation}}

\def\a{\alpha}
\def\b{\beta}

\def\t{\tau}

\def\v{\nu}

\def\m{\mu}
\def\s{\sigma}

\def\o{\omega}

\def\pa{\partial}

\def\e{\epsilon}
\def\G{\Gamma}
\def\md{\mathcal{D}}

\def\d{\delta}

\def\6{\partial}

\def\os{\o^{\mathbf{7}}}
\def\ot{\o^{\mathbf{21}}}
\def\oy{\o^{\mathbf{35}}}
\def\ja{J^{Aj}_{\;\;\;\;\;i}}
\def\mbb{\mathbb{R}}
\newcommand{\een}{\end{equation}}

\newcommand{\ip}{\raise1pt\hbox{\large$\lrcorner$}\,}
\newcommand{\bea}{\begin{eqnarray}}
\newcommand{\eea}{\end{eqnarray}}
\newcommand{\nn}{\nonumber \\}

\begin{document}
\title{The geometry of extended null supersymmetry in M-theory}

\author{Ois\'{\i}n A. P. Mac
  Conamhna\thanks{O.A.P.MacConamhna@damtp.cam.ac.uk} \\ DAMTP \\ Centre
  for Mathematical Sciences \\ University of Cambridge \\ Wilberforce
  Road, Cambridge CB3 0WA, UK.}

\maketitle

\abstract{For supersymmetric spacetimes in eleven dimensions admitting
  a null Killing spinor, a set of explicit necessary and sufficient
  conditions for the existence of any number of arbitrary additional
  Killing spinors is derived. The necessary and sufficient conditions
  are comprised of algebraic relationships, linear in the spinorial
  components, between the spinorial components and their first
  derivatives, and the components of the spin connection and
  four-form. The integrability conditions for the Killing spinor
  equation are also analysed in detail, to determine which components
  of the field equations are implied by arbitrary additional
  supersymmetries and the four-form Bianchi identity. This provides a
  complete formalism for the systematic and exhaustive investigation
  of all spacetimes with extended null supersymmetry in eleven
  dimensions. The formalism is employed to show
  that the general bosonic solution of eleven dimensional supergravity
  admitting a $G_2$ structure defined by four Killing spinors is
  either locally the direct product of $\mathbb{R}^{1,3}$ with a
  seven-manifold of $G_2$ holonomy, or locally the Freund-Rubin direct
  product of $AdS_4$ with a seven-manifold of weak $G_2$ holonomy. In
  addition, all supersymmetric spacetimes admitting a
  $(G_2\ltimes\mathbb{R}^7)\times\mathbb{R}^2$ structure are classified.}   
\clearpage

\section{Introduction}
The problem of systematically classifying supersymmetric spacetimes in
string and M-theory has received great attention in recent years. A
particularly important contribution was that of \cite{gmpw},
which first identified the the usefulness of the concept of
G-structure in this context. Since then, G-structures have been used
for the classification of all (minimally) supersymmetric spacetimes in
many different supergravities, and special supersymmetric
spacetimes of particular physical interest (for example, $AdS$
spacetimes) in string and M-theory; for example,
\cite{gaunt8}-\cite{behr3}. This programme has already led to numerous
successes; for example, to
the discovery of an infinite class of Einstein Sasaki manifolds
\cite{einstein}, together with their field theory duals \cite{j5}, \cite{j6}; a
family of 1/2 BPS excitations of $AdS$ \cite{malda}; supersymmetric
$AdS$ black holes \cite{harvey}, \cite{harvey1}; and supersymmetric
black rings \cite{elvang}-\cite{rings}. In addition, a classification of
all minimally supersymmetric solutions of eleven dimensional
supergravity was given in \cite{gaunt1}, \cite{gaunt}. The systematic and general nature
of this method makes it a natural formalism to extend to the
classification of supersymmetric spacetimes with extended
supersymmetry. Indeed, following the suggestion of \cite{gaunt1}, a
  systematic procedure for applying G-structure ideas to the
  classification of spacetimes with extended supersymmetry was first
  given in \cite{7d}, and illustrated in the context of a seven
  dimensional supergravity. An obvious target for this refined
  G-structure formalism is eleven dimensional supergravity. 

Spinors in eleven dimensions fall into two distinct
categories, according to whether their associated vector is
timelike or null. Equivalently, they may be distinguished by their
isotropy group; a timelike spinor is stabilised by $SU(5)$, while a
null spinor has isotropy group
$(Spin(7)\ltimes\mbb^8)\times\mbb$. Hence the existence of a timelike
or null Killing
spinor implies that the spacetime admits an $SU(5)$ or
$(Spin(7)\ltimes\mbb^8)\times\mbb$ structure respectively\footnote{More
  precisely, a Killing spinor is either null everywhere or it is
  not. In the former case, the spacetime admits a globally defined
  $(Spin(7)\ltimes\mbb^8)\times\mbb$ structure. In the latter case,
  there exists a point and hence a neighbourhood of that point where
  the Killing spinor is timelike, and where it thus defines a
  preferred local $SU(5)$ structure.}. The existence of additional Killing
spinors implies that the structure group is reduced to some subgroup -
the common isotropy group of all the Killing spinors. Thus
supersymmetric spacetimes in eleven dimensions may be naturally split
into two classes: those with structure groups embedding in
$(Spin(7)\ltimes\mbb^8)\times\mbb$, and those with structure groups
embedding in $SU(5)$; of course there is some overlap between these
classes, as spacetimes can admit both timelike and null Killing
spinors. A complete list of the possible structure groups arising as
subgroups of $(Spin(7)\ltimes\mbb^8)\times\mbb$, together with the
spaces of spinors they fix, is given in \cite{nullstructure}.

In \cite{gaunt}, necessary and sufficient conditions on the spin
connection and the four-form for the existence of a single null
Killing spinor were derived. This paper is the fourth and final in a
series \cite{nullstructure}, \cite{spin7}, \cite{eightman}, building on and refining the work of \cite{gaunt}, in
which the G-structure methods of \cite{7d} are used to derive,
given the existence of a single null Killing spinor,
necessary and sufficient conditions for the existence of any number of
arbitrary additional Killing spinors. In \cite{spin7}, the general
solution of the Killing spinor equation admiting a Spin(7) structure
was derived. In \cite{eightman}, spacetimes admitting structure groups
with a compact factor acting non-trivially on eight dimensions were
studied. The results of this series,
together with those of \cite{gaunt}, provide the geometrical ``DNA'' of
all eleven dimensional supersymmetric spacetimes with structure groups
embedding in $(Spin(7)\ltimes\mbb^8)\times\mbb$. The main aim of this
paper is to complete this project, and to provide a complete reference
on extended null supersymmetry in M-theory.

Given the existence of a single null Killing spinor $\e$ in eleven
dimensions, one may always choose the spacetime basis
\be
ds^2=2e^+e^-+\d_{ij}e^ie^j+(e^9)^2,
\end{equation}
where $i,j=1,...,8$, so that the spinor satisfies the projections
\bea
\G_{1234}\e=\G_{3456}\e=\G_{5678}\e=\G_{1357}\e&=&-\e,\nn
\G^+\e&=&0. 
\eea 
The eight-manifold spanned by the $e^i$ is referred to as the
base. With this choice of null Killing spinor, the Spin(7) invariant
four-form defined on the base is given by
\bea
-\phi&=&e^{1234}+e^{1256}+e^{1278}+e^{3456}+e^{3478}+e^{5678}+e^{1357}\nn&-&e^{1368}-e^{1458}-e^{1467}-e^{2358}-e^{2367}-e^{2457}+e^{2468}.
\eea
We
adopt all the conventions of \cite{gaunt}, which are used consistently
throughout this series of papers. Our objective is to extract, from
the Killing spinor equation, the necessary and sufficient conditions
for the existence of an arbitrary additional Killing spinor. Then,
multi-spinor ans\"{a}tze for the Killing spinor equation consistent
with any desired structure group can be made at will. Let us briefly
discuss the method we use; full details are to be found in \cite{7d}-\cite{eightman}.

We construct a basis for the space of Majorana spinors in eleven
dimensions by acting on the Killing spinor equation with a particular
subset of the Clifford algebra. The essential feature we require of
our spinorial basis is that it preserve manifest Spin(7) covariance.
In other words, we decompose the space of Majorana spinors into
modules of the Spin(7) factor of the structure group defined by the
null Killing spinor, and choose an appropriate basis for each
module. Then, by likewise decomposing the spin connection and
four-form into modules of Spin(7), we will see that we may convert
the Killing spinor equation for an arbitrary additional Killing spinor
into a set of purely bosonic equations for the Spin(7) tensors
defining the Killing spinor. Specifically, we choose the basis to
be
\be\label{basiss}
\e,\;\;\G^i\e,\;\;\frac{1}{8}J^A_{ij}\G^{ij}\e,\;\;\G^-\e,\;\;\G^{-i}\e,\;\;\frac{1}{8}J^A_{ij}\G^{-ij}\e,
\een
where the $J^A$, $A=1,...,7$, are a set of two-forms defined on the
base, which furnish a basis for the $\mathbf{7}$ of Spin(7), and we
recognise the decomposition
$\mathbf{32}\rightarrow\mathbf{1}+\mathbf{8}+\mathbf{7}+\mathbf{1^{\prime}}+\mathbf{8^{\prime}}+\mathbf{7^{\prime}}$
of the space of Majorana spinors in eleven dimensions under Spin(7). We choose
the explicit representation of the $J^A$ to be
\bea
J^1=e^{18}+e^{27}-e^{36}-e^{45},& &J^2=e^{28}-e^{17}-e^{35}+e^{46},\nn
J^{3}=e^{38}+e^{47}+e^{16}+e^{25},&&J^4=e^{48}-e^{37}+e^{15}-e^{26}\nn
J^{5}=e^{58}+e^{67}-e^{14}-e^{23},&&J^6=e^{68}-e^{57}-e^{13}+e^{24},\nn
J^7=e^{78}+e^{56}&+&e^{34}+e^{12}.
\eea 
The $J^A$ obey
\be
J^A_{ik}J^{Bk}_{\;\;\;\;\;j}=-\d^{AB}\d_{ij}+K^{AB}_{ij},
\een
where the $K^{AB}_{ij}$ are antisymmetric on $AB$ and $ij$, and
furnish a basis for the $\mathbf{21}$ of Spin(7).

Now, any additional Killing spinor $\eta$ may be written as
\be
\eta=(f+u_i\G^i+\frac{1}{8}f^AJ^A_{ij}\G^{ij}+g\G^-+v_i\G^{-i}+\frac{1}{8}g^AJ^A_{ij}\G^{-ij})\e,
\een
for thirty-two real functions $f,u_i,f^A,g,v_i,g^A$. We may always
simplify some of the additional Killing spinors, by acting on them
with the $(Spin(7)\ltimes\mbb^8)\times\mbb$ isotropy group of $\e$. Since $\e$ is
Killing, $\eta$ is Killing if and only if
\be\label{pr} 
[\md_{\m},f+u_i\G^i+\frac{1}{8}f^AJ^A_{ij}\G^{ij}+g\G^-+v_i\G^{-i}+\frac{1}{8}g^AJ^A_{ij}\G^{-ij}]\e=0,
\een
where
\be
\md_{\m}=\pa_{\m}+\frac{1}{4}\o_{\m\v\s}\G^{\v\s}+\frac{1}{288}(\G_{\m\v\s\t\rho}-8g_{\m\v}\G_{\s\t\rho})F^{\v\s\t\rho}
\een
is the supercovariant derivative. By imposing the projections
satisfied by $\e$, every spacetime component of (\ref{pr}) may be
reduced to a manifest sum of basis spinors; then, the coefficient of
each must vanish separately. Implementing this procedure produces a
set of algebraic relations, linear in the functions defining the
additional Killing spinor, between the functions, their first
derivatives, and the components of the spin connection and fluxes. 

While it is straightforward, the calculation of the commutator is
extremely long and technical. The terms
\be 
[\md_{\m},f+\frac{1}{8}f^AJ^A_{ij}\G^{ij}+g\G^-+\frac{1}{8}g^AJ^A_{ij}\G^{-ij}]\e=0 
\een
have been calculated previously, in \cite{spin7}, \cite{eightman}. Here we complete the calculation, and
impose the $N=1$ constraints \footnote{Here, and throughout, $N$ denotes the number
  of real linearly independent solutions of the $d=11$ Killing spinor
  equation.} on the result, so that it may be used
directly for the analysis of extended supersymmetry. In quoting the
result, we decompose the spin connection and four-form into
irreducible Spin(7) modules; the reader is referred to \cite{gaunt}
for full details of these decompositions.
  
Very recently, in \cite{pap3}, the action of the supercovariant derivative on a
different basis of spinors was computed. In contrast to the present
treatment, the authors of \cite{pap3} did not assume the existence of a
single Killing spinor to begin with, and so do not impose the
conditions for $N=1$ supersymmetry on their result. While this has the
advantage that their results can in principle be used to analyse both
extended timelike and null supersymmetry, it has the disadvantage that
it does not exploit the drastic simplification arising from imposing
the $N=1$ conditions in either case, and may not be used as easily
for the study of extended supersymmetry.

This paper is organised as follows. In section two, we give the full
set of equations obtained from calculating the commutator (\ref{pr}) for a
general additional Killing spinor. We examine the integrability
conditions for the Killing spinor equation to determine which
components of the field equations are implied by the Bianchi identity
and the existence of an arbitrary additional Killing spinor. 

In section three, we use and illustrate the machinery of section two,
to derive the general local bosonic solution of eleven dimensional supergravity,
admitting a $G_2$ structure defined by four Killing spinors. It is
shown that the existence of an $N=4$ $G_2$ structure implies that
spacetime is either locally the direct product of $\mbb^{1,3}$ with a
manifold of $G_2$ holonomy, or locally the direct product of $AdS_4$
with a manifold of weak $G_2$ holonomy. This is derived without making any ansatz
  at all for the bosonic fields, and demanding only that the Killing
  spinors span a subspace of the space of spinors fixed by a $G_2$
  subgroup of Spin(1,10). In section four, all supersymmetric
  spacetimes admitting a $(G_2\ltimes\mbb^7)\times\mbb^2$ structure
  are classified. Section four concludes, and
  some technical material is relegated to the appendix.

\section{Extended null supersymmetry in eleven dimensions}
In this section, we will give the complete equations for arbitrary extended
supersymmetry of a background in M-theory admitting a null Killing
spinor. As discussed in the introduction, these equations are obtained
by imposing
\be\label{ayay}
[\md_{\m},f+u_i\G^i+\frac{1}{8}f^AJ^A_{ij}\G^{ij}+g\G^-+v_i\G^{-i}+\frac{1}{8}g^AJ^A_{ij}\G^{-ij}]\e=0.
\end{equation}
We also impose the $N=1$ conditions on the spin connection and fluxes
on the result we obtain, so that it may be used directly for the
analysis of spacetimes with extended supersymmetry. The $N=1$ conditions
associated with the existence of the Killing spinor $\e$ were derived
in \cite{gaunt}. In order to make this section more self-contained,
we will quote the result of these authors. The constraints on the spin
connection are
\bea
\o_{(\m\v)-}=\os_{ij-}=\os_{-ij}&=&\o_{i9-}=\o_{-9i}=0,\nn
\o_{+9-}&=&\frac{1}{4}\o^{i}_{\;\;i9},\nn
\os_{9ij}&=&-\os_{ij9},\nn
(\os_{[ijk]})^{\mathbf{8}}&=&\frac{1}{56}\phi_{ijk}^{\;\;\;\;\;l}(\o_{99l}-6\o_{l-+}).
\eea
The conditions on the four-form are
\bea
F_{+-9i}&=&2\o_{i-+}-\o_{99i},\nn
F_{+-ij}&=&2\o_{[ij]9},\nn
F_{+9ij}^{\mathbf{7}}&=&2\o_{+ij}^{\mathbf{7}},\nn
F^{\mathbf{8}}_{+ijk}&=&\frac{2}{7}\phi_{ijk}^{\;\;\;\;\;l}\o_{+9l},\nn
F^{\mathbf{7}}_{-9ij}&=&0,\nn
F^{\mathbf{21}}_{-9ij}&=&2\ot_{ij-},\nn
F_{-ijk}&=&0,\nn
F_{9ijk}^{\mathbf{8}}&=&\frac{2}{7}\phi_{ijk}^{\;\;\;\;\;l}(\o_{99l}+\o_{l-+}),\nonumber
\eea
\bea
F^{\mathbf{48}}_{9ijk}&=&-12(\os_{[ijk]})^{\mathbf{48}},\nn
F^{\mathbf{1}}_{ijkl}&=&\frac{3}{7}\o_{+9-}\phi_{ijkl},\nn
F^{\mathbf{7}}_{ijkl}&=&2\phi_{[ijk}^{\;\;\;\;\;m}\os_{l]m9},\nn\label{flux}
F^{\mathbf{35}}_{ijkl}&=&2\phi_{[ijk}^{\;\;\;\;\;m}\oy_{l]m9}.
\eea
Here superscripts refer to irreducible Spin(7) modules; $\os_{ij-}$,
$\os_{ij9}$ denote the $\mathbf{7}$ projections of $\o_{[ij]-}$,
$\o_{[ij]9}$ respectively, and $\ot_{ij-}$ denotes the $\mathbf{21}$
projection of 
$\o_{[ij]-}$. Here and
henceforth, $\o_{ijk}^{\mathbf{7},\mathbf{21}}$ will denote the
$\mathbf{7},\mathbf{21}$ projections of $\o_{ijk}$ on
$j,k$. $(\os_{[ijk]})^{\mathbf{8},\mathbf{48}}$ denote the
$\mathbf{8},\mathbf{48}$ pieces of the completely antisymmetric part
of $\os_{ijk}$. Finally,
$\o_{ij9}^{\mathbf{35}}=\o_{(ij)9}-\frac{1}{8}\d_{ij}\o^k_{\;\;k9}$.
The $F_{+ijk}^{\mathbf{48}}$, $F_{+9ij}^{\mathbf{21}}$ and
$F_{ijkl}^{\mathbf{27}}$ components of the four-form drop out of the
Killing spinor equation for $\e$ and are unconstrained by the $N=1$
constraints. Now, given these constraints, we give the complete set of
conditions for extended supersymmetry. 

\subsection{The complete conditions for extended supersymmetry}
As we have discussed, the equations for extended supersymmetry are derived by imposing
(\ref{ayay}). Wherever
possible, we have eliminated the four-form in favour 
of the spin connection, using the $N=1$ conditions. This is to
maximise the ease with with the conditions for extended supersymmetry
may be used; once the additional conditions on the spin connection
have been solved, the components of the four-form fixed by the $N=1$
constraints may be read off from (\ref{flux}). Essentially, this
procedure produces a set of linear first order partial
differential equations for the functions defining the additional Killing
spinors, which are satisfied by all spacetimes with extended
supersymmetry in eleven dimensions which admit a null Killing spinor;
it extracts the linearly independent
equations contained in the Killing spinor equation. These necessary
and sufficient conditions for arbitrary extended supersymmetry are as follows. \\\\The differential equations for the function $f$ are:\\
\be\label{d-f}
\pa_-f+g\Big[\o_{+9-}-\o_{-+9}\Big]+v^i\Big[\o_{i-+}-\o_{-+i}\Big]=0.
\een\\
\be\label{d+f}
\pa_+f-g\o_{++9}-v^i\o_{++i}=0.
\een\\
\be\label{d9f}
\pa_9f+\frac{1}{3}u^i\Big[\o_{99i}-2\o_{i-+}\Big]+\frac{1}{3}f^A\o_{ij9}J^{Aij}+g\o_{99+}+v^i\Big[\o_{9i+}-\frac{1}{3}\o_{+9i}\Big]-\frac{1}{3}g^A\o_{+ij}J^{Aij}=0.
\een\\
\bea
&&\pa_if+u^j\Big[\frac{1}{2}\d_{ij}\o_{+9-}+\frac{2}{3}\os_{ij9}-\frac{2}{3}\ot_{ij9}-\frac{1}{3}\oy_{ij9}\Big]-f^A\Big[\frac{4}{21}(\o_{j-+}+\o_{99j})\ja+(\os_{[ijk]})^{\mathbf{48}}J^{Ajk}\Big]
\nn&&+g\Big[-\o_{i+9}+\frac{1}{3}\o_{+9i}\Big]+v^j\Big[\o_{ij+}-\frac{5}{3}\os_{+ij}-\frac{1}{6}F^{\mathbf{21}}_{+9ij}\Big]-g^A\Big[\frac{10}{21}\o_{+9j}\ja+\frac{1}{12}F^{\mathbf{48}}_{+ijk}J^{Ajk}\Big]=0.\nn&&\label{dif}
\eea\\
The differential equations for the $u_i$ are:\\
\bea
&&\pa_-u_i+u^j\Big[\o_{-ij}^{\mathbf{21}}+\frac{1}{3}\o_{ij-}^{\mathbf{21}}\Big]+\frac{g}{3}\Big[2\o_{99i}-\o_{i-+}-3\o_{-+i}\Big]+v^j\Big[\o_{-+9}\d_{ij}-\frac{4}{3}\o_{ij9}^{\mathbf{7}}
\nn\label{d-u}&&+\frac{4}{3}\o_{ij9}^{\mathbf{21}}-\frac{2}{3}\o_{ij9}^{\mathbf{35}}\Big]+g^A\Big[(-\o_{-+j}-\frac{5}{7}\o_{j-+}+\frac{2}{7}\o_{99j})J^{Aj}_{\;\;\;\;\;i}-2(\o_{[ijk]}^{\mathbf{7}})^{\mathbf{48}}J^{Ajk}\Big]=0.
\eea\\
\bea
&&\pa_+u_i+u^j\Big[-2\o_{+ij}^{\mathbf{7}}+\o_{+ij}^{\mathbf{21}}+\frac{1}{2}F^{\mathbf{21}}_{+9ij}\Big]+f^A\Big[-\frac{4}{7}\o_{+9j}J^{Aj}_{\;\;\;\;\;i}+\frac{1}{4}F^{\mathbf{48}}_{+ijk}J^{Ajk}\Big]-g\o_{++i}+v_i\o_{++9}
\nn\label{d+u}&&-g^A\o_{++j}J^{Aj}_{\;\;\;\;\;i}=0.
\eea\\
\bea
&&\pa_9u_i+u^j\Big[2\o_{ij9}^{\mathbf{7}}-\ot_{ij9}+\ot_{9ij}-(\frac{1}{2}\o_{+9-}\d_{ij}+\frac{2}{3}\os_{ij9}-\frac{2}{3}\ot_{ij9}-\frac{1}{3}\oy_{ij9})\Big]
\nn&&-f^A\Big[\frac{8}{21}(\o_{j-+}+\o_{99j})J^{Aj}_{\;\;\;\;\;i}
+2(\os_{[ijk]})^{\mathbf{48}}J^{Ajk}\Big]-\frac{g}{3}\Big[3\o_{9+i}+\o_{+9i}\Big]
\nn&&+v^j\Big[-\o_{99+}\d_{ij}-\frac{4}{3}\os_{+ij}+\frac{2}{3}F^{\mathbf{21}}_{+9ij}\Big]
-g^A\Big[(\o_{9+j}-\frac{1}{21}\o_{+9j})J^{Aj}_{\;\;\;\;\;i}
+\frac{1}{6}F^{\mathbf{48}}_{+ijk}J^{Ajk}\Big]=0.\nn&&\label{d9u}
\eea
\bea
&&\pa_iu_j+u^k\Big[\frac{8}{7}\d_{k[i}\o_{j]-+}-(3\os_{ijk}-\ot_{ijk})+\frac{5}{14}\d_{ik}(\o_{99j}-2\o_{j-+})+\frac{1}{7}\d_{jk}(\o_{99i}-2\o_{i-+})
\nn&&-\frac{1}{6}\d_{ij}(\o_{99k}-2\o_{k-+})+4(\os_{[ijk]})^{\mathbf{48}}+\phi_{ij}^{\;\;\;\;lm}(\os_{[klm]})^{\mathbf{48}}\Big]+f^A\Big[\frac{2}{7}\o_{+9-}J^A_{ij}+\frac{1}{12}F^{\mathbf{27}}_{ijkl}J^{Akl}
\nn&&-\frac{1}{6}\d_{ij}\o_{kl9}J^{Akl}-\frac{2}{3}J^{Ak}_{\;\;\;\;\;[i}\ot_{j]k9}
+J^{Ak}_{\;\;\;\;\;(i}\oy_{j)k9}\Big]+g\Big[\o_{ij+}-\frac{1}{3}\os_{+ij}+\frac{1}{2}F^{\mathbf{21}}_{+9ij}\Big]+v^k\Big[\Big(\o_{i+9}
\nn&&+\frac{1}{21}\o_{+9i}\Big)\d_{jk}-\frac{10}{21}\d_{ik}\o_{+9j}+\frac{2}{21}\d_{ij}\o_{+9k}-\frac{2}{21}\phi_{ijk}^{\;\;\;\;\;\;l}\o_{+9l}+\frac{2}{3}F^{\mathbf{48}}_{+ijk}+\frac{1}{4}F^{\mathbf{48}}_{+jlm}\phi_{ik}^{\;\;\;\;lm}
\nn&&-\frac{1}{12}F^{\mathbf{48}}_{+klm}\phi_{ij}^{\;\;\;\;lm}\Big]+g^A\Big[\frac{1}{6}\d_{ij}\o_{+kl}J^{Akl}+(\o_{ik+}+\os_{+ik}+\frac{1}{6}F^{\mathbf{21}}_{+9ik})J^{Ak}_{\;\;\;\;\;j}+\frac{1}{3}F^{\mathbf{21}}_{+9jk}J^{Ak}_{\;\;\;\;\;i}\Big]=0.
\nn&&\label{diu}
\eea
The differential equations for the $f^A$ are:\\
\bea
&&\pa_-f^A+\frac{1}{4}f^B\Big[\o_{-ij}-\frac{1}{3}\o_{ij-}\Big]K^{BAij}+\frac{2}{3}g\o_{ij9}J^{Aij}+v^i\Big[\Big(\frac{13}{21}\o_{j-+}-\o_{-+j}-\frac{8}{21}\o_{99j}\Big)\ja
\nn&&-2(\os_{[ijk]})^{\mathbf{48}}J^{Ajk}\Big]+g^B\Big[-(\o_{-+9}+\frac{1}{7}\o_{+9-})\d^{AB}-\frac{1}{24}J^{Aij}F^{\mathbf{27}}_{ijkl}J^{Bkl}+\frac{1}{3}\o_{ij9}K^{BAij}\Big]=0.\nn&&\label{d-fA}
\eea
\be\label{d+fA}
\pa_+f^A+u^i\Big[\frac{4}{7}\o_{+9j}\ja-\frac{1}{4}F^{\mathbf{48}}_{+ijk}J^{Ajk}\Big]+\frac{1}{4}f^B\Big[\o_{+ij}-\frac{1}{2}F_{+9ij}\Big]K^{BAij}-v^i\o_{++j}\ja-g^A\o_{++9}=0.
\een
\bea
&&\pa_9f^A+u^i\Big[\Big(\frac{5}{7}\o_{99j}-\frac{2}{7}\o_{j-+}\Big)\ja+2(\os_{[ijk]})^{\mathbf{48}}J^{Ajk}\Big]+f^B\Big[\frac{1}{4}(\o_{9ij}+\o_{ij9})K^{BAij}
\nn&&-\Big(\frac{4}{7}\d^{AB}\o_{+9-}+\frac{1}{48}J^{Aij}F^{\mathbf{27}}_{ijkl}J^{Bkl}+\frac{1}{6}\o_{ij9}K^{BAij}\Big)\Big]+\frac{1}{3}g\o_{+ij}J^{Aij}-v^i\Big[(\o_{9+j}-\frac{1}{21}\o_{+9j})\ja
\nn\label{d9fA}&&+\frac{1}{6}F^{\mathbf{48}}_{+ijk}J^{Ajk}\Big]+g^B\Big[\d^{AB}\o_{99+}+\frac{1}{6}F_{+9ij}K^{BAij}\Big]=0.
\eea\\
\bea
&&\pa_if^A+u^j\Big[-\frac{3}{28}\o_{+9-}J^A_{ij}-\frac{1}{4}F^{\mathbf{27}}_{ijkl}J^{Akl}+\frac{1}{2}\d_{ij}J^{Akl}\os_{kl9}+\os_{jk9}J^{Ak}_{\;\;\;\;\;i}-(\ot_{ik9}+\frac{1}{2}\oy_{ik9})J^{Ak}_{\;\;\;\;\;j}
\nn&&+\Big(\frac{1}{4}\d_{jk}\o_{+9-}-\frac{1}{3}\os_{jk9}+\frac{1}{3}\ot_{jk9}-\frac{1}{6}\oy_{jk9}\Big)J^{Ak}_{\;\;\;\;\;i}\Big]
+f^B\Big[-\frac{1}{2}J^{[B}_{ij}J^{A]}_{kl}\o^{jkl}+\frac{1}{4}\Big(\o_{ijk}K^{BAjk}
\nn&&-(2\o_{j-+}+\o_{99j})K^{BAj}_{\;\;\;\;\;\;\;\:i}\Big)+\Big(\frac{4}{21}(\o_{k-+}+\o_{99k})J^{Bk}_{\;\;\;\;\;j}+(\os_{[jkl]})^{\mathbf{48}}J^{Bkl}\Big)\ja\Big]+g\Big[\frac{2}{21}\o_{+9j}\ja
\nn&&+\frac{1}{4}F^{\mathbf{48}}_{+ijk}J^{Ajk}\Big]+v^j\Big[\frac{1}{6}\d_{ij}\os_{+kl}J^{Akl}+(\o_{ik+}-\frac{1}{3}\os_{+ik}-\frac{1}{2}F^{\mathbf{21}}_{+9ik})J^{Ak}_{\;\;\;\;\;j}+\frac{1}{3}F^{\mathbf{21}}_{+9jk}J^{Ak}_{\;\;\;\;\;i}\Big]
\nn&&+g^B\Big[-\d^{AB}(\o_{i+9}+\frac{13}{126}\o_{+9i})-\frac{1}{18}\o_{+9j}K^{BAj}_{\;\;\;\;\;\;\;\:i}-\frac{17}{144}F^{\mathbf{48}}_{+ijk}K^{BAjk}+\frac{1}{12}J^{(A}_{ij}J^{B)}_{kl}F^{\mathbf{48}jkl}_+\Big]=0.\nn&&\label{difA}
\eea
The differential equations for $g$ are:\\
\be\label{d-g}
\pa_-g=0.
\een\\
\be\label{d+g}
\pa_+g+\frac{1}{3}u^i\Big[\o_{99i}-2\o_{i-+}\Big]+\frac{1}{3}f^A\o_{ij9}J^{Aij}-\frac{2}{3}v^i\o_{+i9}-\frac{1}{3}g^A\o_{+ij}J^{Aij}=0.
\een\\
\be\label{d9g}
\pa_9g+v^i\o_{99i}=0.
\een\\
\bea
&&\pa_ig-\frac{2}{3}u^j\ot_{ij-}+\frac{1}{3}g\Big[2\o_{i-+}-\o_{99i}\Big]-v^j\Big[\d_{ij}\o_{+9-}+\frac{10}{3}\os_{ij9}+\frac{2}{3}\ot_{ij9}+\frac{2}{3}\oy_{ij9}\Big]
\nn\label{dig}&&+g^A\Big[\frac{1}{2}(\o_{99j}\ja-\o_{ijk}J^{Ajk})-\Big(\frac{1}{7}(\o_{j-+}+\o_{99j})\ja-(\os_{[ijk]})^{\mathbf{48}}J^{Ajk}\Big)\Big]=0.
\eea\\
The differential equations for the $v^i$ are:\\
\be\label{d-v}
\pa_-v_i+v^j\Big[\ot_{-ij}-\ot_{ij-}\Big]=0.
\een\\
\bea
&&\pa_+v_i+u^j\Big[\frac{1}{2}\d_{ij}\o_{+9-}+\frac{2}{3}\os_{ij9}-\frac{2}{3}\ot_{ij9}-\frac{1}{3}\oy_{ij9}\Big]-f^A\Big[\frac{4}{21}(\o_{j-+}+\o_{99j})\ja+(\os_{[ijk]})^{\mathbf{48}}J^{Ajk}\Big]
\nn\label{d+v}&&-\frac{2}{3}g\o_{+9i}+v^j\Big[-\frac{2}{3}\os_{+ij}+\ot_{+ij}-\frac{1}{6}F^{\mathbf{21}}_{+9ij}\Big]-g^A\Big[\frac{10}{21}\o_{+9j}\ja+\frac{1}{12}F^{\mathbf{48}}_{+ijk}J^{Ajk}\Big]=0.
\eea\\
\bea
&&\pa_9v_i+\frac{2}{3}u^j\ot_{ij-}-\frac{2}{3}g\Big[\o_{99i}+\o_{i-+}\Big]+v^j\Big[\frac{1}{2}\o_{+9-}\d_{ij}+\frac{4}{3}\os_{ij9}+\ot_{9ij}-\frac{1}{3}\ot_{ij9}-\frac{1}{3}\oy_{ij9}\Big]
\nn\label{d9v}&&+g^A\Big[-\frac{2}{7}(\o_{j-+}+\o_{99j})\ja+2(\os_{[ijk]})^{\mathbf{48}}J^{Ajk}\Big]=0.
\eea\\
\bea
&&\pa_iv_j+\frac{2}{3}f^AJ^{Ak}_{\;\;\;\;\;[i}\ot_{j]k-}+g\Big[\frac{1}{2}\d_{ij}\o_{+9-}+\frac{4}{3}\os_{ij9}+\oy_{ij9}\Big]+v^k\Big[\frac{8}{21}(\o_{99[i}+\o_{[i|-+|})\d_{j]k}
\nn&&-\frac{1}{6}\phi_{ijk}^{\;\;\;\;\;\;l}(\o_{99l}-2\o_{l-+})+\o_{ijk}-4(\os_{[ijk]})^{\mathbf{48}}-\phi_{ij}^{\;\;\;\;lm}(\os_{[lmk]})^{\mathbf{48}}\Big]+g^A\Big[\frac{3}{14}\o_{+9-}J^A_{ij}
\nn\label{div}&&-\frac{1}{12}F^{\mathbf{27}}_{ijkl}J^{Akl}
-\frac{2}{3}J^{Ak}_{\;\;\;\;\;[i}\ot_{j]k9}-J^{Ak}_{\;\;\;\;\;[i}\oy_{j]k9}\Big]=0.
\eea\\
The differential equations for the $g^A$ are:\\
\be\label{d-gA}
\pa_-g^A+\frac{1}{4}g^B\Big[\o_{-ij}+\o_{ij-}]K^{BAij}=0.
\een\\
\bea
&&\pa_+g^A+u^i\Big[\Big(\frac{1}{7}\o_{99j}-\frac{6}{7}\o_{j-+}\Big)\ja-(\os_{[ijk]})^{\mathbf{48}}J^{Ajk}\Big]-f^B\Big[\frac{4}{7}\d^{AB}\o_{+9-}+\frac{1}{48}J^{Aij}F^{\mathbf{27}}_{ijkl}J^{Bkl}
\nn&&+\frac{1}{6}\o_{ij9}K^{BAij}\Big]
+\frac{1}{3}g\o_{+ij}J^{Aij}+v^i\Big[\frac{10}{21}\o_{+9j}\ja+\frac{1}{12}F^{\mathbf{48}}_{+ijk}J^{Ajk}\Big]
\nn\label{d+gA}&&+\frac{1}{4}g^B\Big[\o_{+ij}+\frac{1}{6}F_{+9ij}\Big]K^{BAij}=0.
\eea\\
\bea
&&\pa_9g^A+\frac{1}{6}f^B\o_{ij-}K^{BAij}-\frac{1}{3}g\o_{ij9}J^{Aij}+v^i\Big[\Big(\frac{13}{21}\o_{99j}-\frac{8}{21}\o_{j-+}\Big)\ja-2(\os_{[ijk]})^{\mathbf{48}}J^{Ajk}\Big]
\nn\label{d9gA}&&+g^B\Big[\frac{1}{4}(\o_{9ij}+\o_{ij9})K^{BAij}+\Big(\frac{4}{7}\d^{AB}\o_{+9-}+\frac{1}{48}J^{Aij}F^{\mathbf{27}}_{ijkl}J^{Bkl}-\frac{1}{6}\o_{ij9}K^{BAij}\Big)\Big]=0.
\eea\\
\bea
&&\pa_ig^A-u^j\Big[\ot_{ik-}J^{Ak}_{\;\;\;\;\;j}-\frac{1}{3}\ot_{jk-}J^{Ak}_{\;\;\;\;\;i}\Big]+g\Big[-\frac{2}{21}(\o_{99j}+\o_{j-+})\ja+3(\os_{[ijk]})^{\mathbf{48}}J^{Ajk}\Big]
\nn&&+v^j\Big[-\frac{1}{7}\o_{+9-}J^A_{ij}+\frac{1}{4}F^{\mathbf{27}}_{ijkl}J^{Akl}-\frac{1}{3}\d_{ij}\o_{kl9}J^{Akl}-\Big(\frac{4}{3}\os_{ik9}+\ot_{ik9}+\frac{1}{2}\oy_{ik9}\Big)J^{Ak}_{\;\;\;\;\;j}+\Big(-\frac{2}{3}\os_{jk9}
\nn&&+\frac{1}{3}\ot_{jk9}+\frac{1}{6}\oy_{jk9}\Big)J^{Ak}_{\;\;\;\;\;i}\Big]+g^B\Big[\d^{AB}\o_{i-+}+\frac{1}{4}\Big(\o_{ijk}K^{BAjk}+(2\o_{j-+}-\o_{99j})K^{BAj}_{\;\;\;\;\;\;\;\:i}\Big)
\nn\label{digA}&&+\frac{1}{2}J^{[B}_{ij}J^{A]}_{kl}\o^{jkl}+\Big(\frac{1}{7}(\o_{k-+}+\o_{99k})J^{Bk}_{\;\;\;\;\;j}-(\os_{[jkl]})^{\mathbf{48}}J^{Bkl}\Big)\ja\Big]=0.
\eea

\subsection{Integrability conditions}\label{oko}
The existence of a solution of the Killing spinor equation implies
that some components of the field equations and Bianchi identity are
identically satisfied. This follows from the (contracted)
integrability condition for the Killing spinor equation. We will not
undertake a complete analysis of the integrability condition here
(though there it is entirely straightforward to do so, it comes at a
significant computational cost);
rather, we will assume that the Bianchi identity for the four-form,
$dF=0$, is
always imposed on the solution of the Killing spinor equation. We will
then use the integrability condition to determine which components of
the field equations must be imposed on the solution of the Killing
spinor equation and Bianchi identity. The Einstein equation is
\be
R_{\m\v}-\frac{1}{12}\Big(F_{\m\s\t\rho}F_{\v}^{\;\;\s\t\rho}-\frac{1}{12}g_{\m\v}F_{\s\t\rho\lambda}F^{\s\t\rho\lambda})=0.
\een
The (classical) four-form field equation is
\be
\star\Big(d\star F+\frac{1}{2}F\wedge F\Big)=0.
\een
Note that in full M-theory both the field equations and the supersymmetry
transformation receive higher-order corrections; while it is
conceptually straightforward to incorporate these corrections into the
formalism, here we restrict
attention to classical supergravity. 
Given that the Bianchi identity is imposed, the contracted
integrability condition for an arbitrary Killing spinor $\eta$ reads
\be\label{inttt}
\G^{\v}[\md_{\m},\md_{\v}]\eta=(E_{\m\v}\G^{\v}+Q^{\v\s\t}\G_{\m\v\s\t}-6Q_{\m\v\s}\G^{\v\s})\eta=0,
\end{equation}
where up to constant overall factors the Einstein and four-form field equations are respectively
$E_{\m\v}=0$, $Q_{\m\v\s}=0$. By taking $\eta=\e$, and writing each
spacetime component of (\ref{inttt}) as a manifest sum of basis spinors,
we find the following algebraic relationships between the components
of the field equations:
\bea
E_{+-}=E_{99}&=&12Q_{+-9},\nn E_{+i}&=&18Q_{+i9},\nn
E_{ij}&=&-6Q_{+-9}\d_{ij}.
\eea
The components $E_{++}$ and $Q_{+ij}^{\mathbf{21}}$ drop out of the
integrability condition for $\e$ and are unconstrained. All other
components of the field equations vanish identically. Thus, given the
existence of the Killing spinor $\e$ and that the Bianchi identity is
satisfied, it is sufficient to impose
$E_{++}=Q_{+9i}=Q_{+-9}=Q_{+ij}^{\mathbf{21}}=0$ to ensure that all
field equations are satisfied.

Given the existence of a Killing spinor $\e$ and that the Bianchi
identity is satisfied, the integrability condition for an arbitrary
additional Killing spinor is
\be
[\G^{\v}[\md_{\m},\md_{\v}],f+u_i\G^i+\frac{1}{8}f^AJ^A_{ij}\G^{ij}+g\G^-+v_i\G^{-i}+\frac{1}{8}g^AJ^A_{ij}\G^{-ij}]\e=0,
\end{equation}  
with $\G^{\v}[\md_{\m},\md_{\v}]$ given by (\ref{inttt}). This
equation in analysed in detail in the appendix, and here we will quote
the result.\\\\If in addition to $\e$ there exists a Killing spinor with
$v_iv^i\neq0$, then it is sufficient to impose the Bianchi identity
to ensure that all field equations are satisfied.\\\\If in addition
to $\e$ there exists
a Killing spinor with $v_i=0$, $g^2+g^Ag^A\neq0$, then it
is sufficient to impose the Bianchi identity and $Q_{+-9}=0$ to ensure
that all field equations are satisfied.\\\\If
in addition to $\e$ there exists a Killing spinor with $v_i=g=g^A=0$,
$u_iu^i\neq0$, then it is sufficient to impose the Bianchi identity
and $E_{++}=Q_{+ij}^{\mathbf{21}}=0$ to ensure that all field equations are
satisfied.\\\\If in addition to $\e$ there exists a Killing spinor
with $v_i=g=g^A=u_i=0$, $f^Af^A\neq0$, then it is sufficient to impose
the Bianchi identity and $E_{++}=Q_{+-9}=Q_{+9i}=Q_{+ij}^{\mathbf{21}}=0$ to ensure
that all field equations are satisfied.

\section{The general solution of eleven dimensional supergravity admitting an $N=4$ $G_2$ structure}
In this section, we will illustrate how the full machinery of the
previous section may be employed to perform exhaustive classifications
of supersymmetric spacetimes. Specifically, we will derive the general
local bosonic solution of eleven dimensional supergravity which admits four
Killing spinors defining a $G_2$ structure. As was shown in
\cite{nullstructure}, additional Killing spinors defining a $G_2$ structure may be taken to
be of the form
\be\label{g2}
(f+u_8\G^8+g\G^-+v_8\G^{-8})\e.
\een
We demand that in addition to $\e$, there exist three linearly
independent solutions of the Killing spinor of the form (\ref{g2}). We
will thus classify all spacetimes with maximal $G_2$ supersymmetry,
since four is the greatest number of Killing spinors compatible with a
$G_2$ structure. 

As was pointed out in \cite{7d}, \cite{eightman}, supersymmetric spacetimes
admitting the maximal number of Killing spinors consistent with a
particular structure group are particularly easy to classify. This is
because of the following argument. The conditions for supersymmetry
given in the previous section are a set of $11\times32$ partial
differential equations, one for each spacetime component of the
partial derivative of each spinorial component. For the case of
maximal $G_2$ supersymmetry, by demanding the existence of three
additional Killing spinors of the specific form (\ref{g2}), we are
assuming that the spinors span a four dimensional subspace of the
space of spinors. Thus the full set of equations for each Killing
spinor becomes a set of $11\times 4$ differential equations for the
spinorial components, together with a set of $11\times 28$ algebraic
equations for the spinorial components. Each of the algebraic
equations for the spinorial components is of the schematic form
\be     
u_8A+gB+v_8C=0.
\een
The specific property of maximal $G_2$ supersymmetry which greatly
facilitates solving the Killing spinor equation, is that since we are
demanding the existence of three additional linearly independent
Killing spinors, in
each of the algebraic equations for the spinorial components we must
have $A=B=C=0$. Similar arguments apply to
maximal supersymmetry consistent with any other structure group. Let
us now work through this procedure in detail.

\subsection{The conditions for maximal $G_2$ supersymmetry}
The $G_2$ structure induces a $1+7$ split of the tangent space of the
eight dimensional base; we have chosen the frame so that the $G_2$
structure group acts non-trivially on the $e^A$, $A=1,..,7$. We have
also chosen the $J^A$, given in the introduction, such that
\bea
J^A_{B8}&=&\d_{AB},\nn
J^A_{BC}&=&-\Phi_{ABC},
\eea
where $\Phi$ is the $G_2$-invariant associative three-form,
\be
\Phi_{ABC}=\phi_{ABC8}.
\een
Let us briefly discuss the decomposition of Spin(8) two-forms into
irreducible modules of Spin(7) and $G_2$. Under Spin(7), the
$\mathbf{28}$ of Spin(8) decomposes as
$\mathbf{28}\rightarrow\mathbf{7}+\mathbf{21}$. Under $G_2$, the
$\mathbf{7}$ of Spin(7) is irreducible, while the $\mathbf{21}$
decomposes as
$\mathbf{21}\rightarrow\mathbf{7^{\prime}}+\mathbf{14}$. We may effect
the decomposition of a two-form $\a$ in the $\mathbf{28}$ of Spin(8) into
irreducible modules of Spin(7) using the $J^A$ and $K^{AB}$:
\bea
\a^{\mathbf{7}}_{ij}&=&\frac{1}{8}\a_{kl}J^{Akl}J^A_{ij},\nn\label{mnmn}
\a^{\mathbf{21}}_{ij}&=&\frac{1}{16}\a_{kl}K^{ABkl}K^{AB}_{ij}.
\eea
We may construct bases for the $\mathbf{7^{\prime}}$ and the
$\mathbf{14}$ by taking appropriate linear combinations of the
$K^{AB}_{ij}$. These linear combinations are obtained by applying
$G_2$ projectors, as follows. A basis for the $\mathbf{7^{\prime}}$ is given by 
\be
K^{AB\mathbf{7^{\prime}}}_{ij}=\frac{1}{3}\Big(K^{AB}_{ij}-\frac{1}{2}\Upsilon^{ABCD}K^{CD}_{ij}\Big),
\een
while a basis for the $\mathbf{14}$ is given by
\be
K^{AB\mathbf{14}}_{ij}=\frac{2}{3}\Big(K^{AB}_{ij}+\frac{1}{4}\Upsilon^{ABCD}K^{CD}_{ij}\Big),
\een
where $\Upsilon$ is the $G_2$ invariant coassociative four-form,
\be
\Upsilon_{ABCD}=\phi_{ABCD},
\een
and $\Upsilon=\star_7\Phi$, with orientation fixed by
\be
\e_{ABCDEFG}=\e_{ABCDEFG8}.
\een 
For a two-form $\beta$ in the
$\mathbf{14}$ of $G_2$,
$\Phi_A^{\;\;BC}\b^{\mathbf{14}}_{BC}=0$. Then, since
$K^{BC}_{A8}=\Phi_{ABC}$, we see from (\ref{mnmn}) that if
\be
\a^{\mathbf{21}}_{A8}=0,
\een
then $\a_{AB}=\a^{\mathbf{14}}_{AB}$.
\\\\
Let us now solve the algebraic equations for the spinorial
components. All three additional Killing spinors have
$u_A=f^A=v_A=g^A=0$, so by 
the argument given above, the coefficients of $u_8$, $g$ and $v_8$
must vanish in all the differential equations for $u_A$, $f^A$, $v_A$
and $g^A$. From the coefficient of $g$ in (\ref{d9gA}), we find
\be
\os_{ij9}=0.
\een
From the coefficient of $v_8$ in the $A$ component of (\ref{d-u}),
\be
\ot_{A89}=\frac{1}{2}\oy_{A89},
\een
and from the coefficient of $u_8$ in the $A$ component of (\ref{d+v}),
\be
\ot_{A89}=-\frac{1}{2}\oy_{A89}.
\een
Hence
\be
\ot_{A89}=\oy_{A89}=0.
\een
From the coefficient of $u_8$ in the $A$ component of (\ref{d9u}),
\be
\ot_{9A8}=0.
\een
From the coefficient of $g$ in the $iA$ component of (\ref{div}),
\be
\oy_{iA9}=-\frac{1}{2}\d_{iA}\o_{+9-},
\een
and hence, since $\oy_{ij9}$ is traceless on $i,j$,
\be
\oy_{889}=\frac{7}{2}\o_{+9-}.
\een
From the coefficient of $v_8$ in (\ref{digA}), 
\be\label{44}
-\frac{1}{7}\o_{+9-}\d_{iA}+\frac{1}{4}F_{i8jk}^{\mathbf{27}}J^{Ajk}-\ot_{iA9}-\frac{1}{2}\oy_{iA9}-\frac{1}{6}\oy_{889}\d_{iA}=0.
\een
The $i=8$ component of this equation produces nothing new. As in \cite{eightman}, we express $F^{\mathbf{27}}$ as
\be
F^{\mathbf{27}}=f^{AB}\Big(J^A\wedge J^B-\frac{1}{7}\d^{AB}J^C\wedge
J^C\Big),
\een
so that
\be
F^{\mathbf{27}}_{B8jk}J^{Ajk}=8\Big(f^{(AB)}-\frac{1}{7}\d^{AB}f^{CC}\Big).
\een   
Then taking $i=B$ in (\ref{44}), from the antisymmetric part on $A,B$
we find
\be
\ot_{AB9}=0.
\een
Tracing on $A,B$ we find
\be
\o_{+9-}=\frac{14}{9}\oy_{889},
\end{equation}
and hence $\o_{+9-}=\oy_{ij9}=0$. Finally, the vanishing of the
symmetric traceless part of (\ref{44}) implies that
\be
F^{\mathbf{27}}_{ijkl}=0.
\een
Using the $N=1$ conditions, we may summarise the additional
constraints we have derived hitherto as
\bea
\o_{+9-}=\o_{ij9}=\o_{9A8}&=&F^{\mathbf{27}}_{ijkl}=0,\nn
\o_{9AB}&=&\o^{\mathbf{14}}_{9AB},
\eea
where $\o_{9AB}^{\mathbf{14}}$ denotes the projection of $\o_{9AB}$ on
the $\mathbf{14}$, or adjoint, of $G_2$. \\\\Next, from the
coefficient of $u_8$ in (\ref{digA}), 
\be
\ot_{iA-}-\frac{1}{3}\ot_{8j-}\ja=0.
\een
Taking $i=8$ gives
\be
\ot_{A8-}=0,
\een
and then combined with the $N=1$ conditions, $i=B$ produces
\be
\o_{ij-}=0.
\een
Then the $A$ component of (\ref{d-v}), together with the $N=1$
conditions, gives
\bea
\o_{-A8}&=&0,\nn
\o_{-AB}&=&\o_{-AB}^{\mathbf{14}}.
\eea\\\\Next, from the coefficient of $g$ in (\ref{difA}),
\be\label{101}
\frac{2}{21}\o_{+9j}\ja+\frac{1}{4}F^{\mathbf{48}}_{+ijk}J^{Ajk}=0.
\een
From the coefficient of $u_8$ in (\ref{d+fA}), 
\be
\frac{4}{7}\o_{+9A}-\frac{1}{4}F^{\mathbf{48}}_{+8jk}J^{Ajk}=0.
\een
Comparing with the $i=8$ component of (\ref{101}), we find
\be
\o_{+9A}=F^{\mathbf{48}}_{+8jk}J^{Ajk}=0.
\een
Then the $i=B$ component of (\ref{101}) reads
\be
-\frac{2}{21}\o_{+98}\d_{AB}-\frac{1}{4}F^{\mathbf{48}}_{+BCD}\Phi^{ACD}-\frac{1}{2}F^{\mathbf{48}}_{+8AB}=0.
\een
Tracing on $A,B$, and using the fact that for any three-form $\a^{\mathbf{48}}$ in the
$\mathbf{48}$ of Spin(7),
\be
\a_{ABC}^{\mathbf{48}}\Phi^{ABC}=\a_{ijk}^{\mathbf{48}}\phi^{ijk}_{\;\;\;\;\;8}=0,
\een
we get
\be
\o_{+9i}=0.
\een
Then $F^{\mathbf{48}}_{+ijk}J^{Ajk}=0$, which implies that
\be
F_{+ijk}^{\mathbf{48}}=0.
\een
Then from the coefficient of $v_8$ in (\ref{d9fA}), we get
\be
\o_{9+A}=0.
\een\\\\Next, from the coefficient of $g$ in (\ref{digA}),
\be\label{102}
-\frac{2}{21}(\o_{99j}+\o_{j-+})\ja+3(\os_{[ijk]})^{\mathbf{48}}J^{Ajk}=0,
\een
From the coefficient on $v_8$ in (\ref{d9gA}), 
\be\label{103}
\frac{13}{21}\o_{99A}-\frac{8}{21}\o_{A-+}-2(\os_{[8jk]})^{\mathbf{48}}J^{Ajk}=0.
\een
From the coefficient of $u_8$ in (\ref{d+gA}),
\be\label{104}
\frac{1}{7}\o_{99A}-\frac{6}{7}\o_{A-+}-(\os_{[8jk]})^{\mathbf{48}}J^{Ajk}=0.
\een 
Combined with the $i=8$ component of (\ref{102}), equations
(\ref{103}) and (\ref{104}) imply that
\be\label{105}
\o_{99A}=\o_{A-+}=(\os_{[8jk]})^{\mathbf{48}}J^{Ajk}=0.
\end{equation}
Then taking $i=A$ in (\ref{102}), we get
\bea\label{106}
\o_{998}&=&-\o_{8-+},\\
(\os_{[ijk]})^{\mathbf{48}}&=&0.
\eea
From the coefficient of $v_8$ in (\ref{d-fA}), 
\be
\o_{-+A}=0.
\een
From the coefficient of $v_8$ in the $iA$ component of (\ref{div}),
\be\label{107}
\o_{iA8}=0.
\een
From the coefficient of $u_8$ in the $iA$ component of (\ref{diu}),
\be\label{108}
3\os_{iA8}-\ot_{iA8}-\frac{1}{2}\d_{iA}\o_{8-+}=0.
\een
From (\ref{105}) and (\ref{106}), together with the $N=1$ conditions
and the expression for the intrinsic contorsion of a Spin(7) structure
given in \cite{gaunt}, we may write
\be\label{109}
\os_{ijk}=\frac{1}{4}\d_{i[j}\o_{k]-+}-\frac{1}{8}\phi_{ijk}^{\;\;\;\;\;l}\o_{l-+}.
\een
Since $\o_{A-+}=0$, $\os_{8jk}=0$, (\ref{107}) implies that
$\ot_{88A}=0$, and hence that
\be
\o_{8AB}=\o^{\mathbf{14}}_{8AB}.
\een
Furthermore, (\ref{107}) also implies that $\os_{AB8}=-\ot_{AB8}$, so
(\ref{108}) is implied by (\ref{107}) and (\ref{109}). Note that the
$i=B$ component of (\ref{108}) may be rewritten as
\be
\Phi_{A}^{\;\;\;CD}\o_{BCD}=-\d_{AB}\o_{8-+}.
\een \\\\Next, from the coefficient of $g$ in (\ref{d+gA}), 
\be
\os_{+ij}=0.
\een
From the coefficient of $v_8$ in the $A$ component of (\ref{d9u}),
\be
F^{\mathbf{21}}_{+9A8}=0.
\een
Then from the coefficient of $v_8$ in the $A$ component of (\ref{d+v}),
\bea
\ot_{+A8}&=&0,\nn
\o_{+AB}&=&\o_{+AB}^{\mathbf{14}}.
\eea
From the coefficient of $v_8$ in (\ref{difA}),
\be
\o_{iA+}=\frac{1}{2}F^{\mathbf{21}}_{+9iA}.
\een
From the coefficient of $g$ in the $iA$ component of (\ref{diu}),
\be
\o_{iA+}=-\frac{1}{2}F^{\mathbf{21}}_{+9iA}.
\een
Thus
\be
F^{\mathbf{21}}_{+9ij}=\o_{iA+}=0.
\een \\\\Finally, from the coefficient of $v_8$ in (\ref{d+fA}), we
find
\be
\o_{++A}=0.
\een
\paragraph{Summary}At this point we have solved all the algebraic equations for
the spinorial components, so let us summarise what we have
found. There are the following conditions on the spin connection:
\bea
\o_{(\m\v)-}=\o_{i9-}&=&\o_{-9i}=0,\nn
\o_{+9-}=\o_{ij9}&=&\o_{9A8}=0,\nn
\o_{9AB}&=&\o_{9AB}^{\mathbf{14}},\nn
\o_{ij-}&=&\o_{-A8}=0,\nn
\o_{-AB}&=&\o_{-AB}^{\mathbf{14}},\nn
\o_{+9i}&=&\o_{9+A}=0,\nn
\o_{99A}=\o_{A-+}&=&\o_{-+A}=0,\nn
\o_{998}&=&-\o_{8-+},\nn
\o_{88A}&=&\o_{AB8}=0,\nn
\o_{8AB}&=&\o_{8AB}^{\mathbf{14}},\nn
\os_{ijk}&=&\frac{1}{4}\d_{i[j}\o_{k]-+}-\frac{1}{8}\phi_{ijk}^{\;\;\;\;\;l}\o_{l-+},
\nn
\o_{+A8}&=&0,\nn
\o_{+AB}&=&\o_{+AB}^{\mathbf{14}},\nn
\o_{8A+}=\o_{AB+}&=&\o_{++A}=0.
\eea
We have also found that the components of the flux not fixed by the
$N=1$ constraints must vanish,
$F_{+ijk}^{\mathbf{48}}=F^{\mathbf{21}}_{+9ij}=F^{\mathbf{27}}_{ijkl}=0$.
Inserting the additional conditions we have derived for the spin
connection into 
the $N=1$ expressions for the flux, we find that the only nonzero
component is
\be
F_{+-98}=3\o_{8-+}.
\een\\\\It remains to solve the differential equations for $f$, $u_8$,
$g$ and $v_8$, for each of the four Killing spinors. Given the
constraints we have derived on the spin connection, these equations
reduce to
\be
\pa_-f-g\o_{-+9}+v_8\Big[\o_{8-+}-\o_{-+8}\Big]=0,
\een
\be
\pa_+f-g\o_{++9}-v_8\o_{++8}=0,
\een
\be
\pa_9f-u_8\o_{8-+}+g\o_{99+}-v_8\o_{9+8}=0,
\een
\be
\pa_if-g\o_{i+9}+v_8\o_{i8+}=0.
\een\\\\\be
\pa_-u_8-g\Big[\o_{8-+}+\o_{-+8}\Big]+v_8\o_{-+9}=0,
\een
\be
\pa_+u_8-g\o_{++8}+v_8\o_{++9}=0,
\een
\be
\pa_9u_8-g\o_{9+8}-v_8\o_{99+}=0,
\een
\be
\pa_iu_8-u_8\d_{i8}\o_{8-+}+g\o_{i8+}+v_8\o_{i+9}=0.
\een\\\\\be
\pa_-g=0,
\een
\be
\pa_+g-u_8\o_{8-+}=0,
\een
\be
\pa_9g-v_8\o_{8-+}=0,
\een
\be
\pa_8g+g\o_{8-+}=0,
\een
\be
\pa_Ag=0.
\een\\\\\be
\pa_{\m}v_8=0.
\een

\subsection{Solving the conditions for maximal $G_2$ supersymmetry}
The only assumption we have made in deriving the above conditions is
that there exist four linearly independent solutions of the Killing
spinor equation stabilised by a common $G_2$ subgroup of
Spin(1,10). Now we must solve the differential equations for the
spinorial components, and also solve the conditions on the spin
connection to determine the metric. To do so, we exploit the fact that
we still have a lot of freedom left in our choice of spacetime
basis. We are free to perform $G_2$ rotations of the $e^A$, leaving all
the Killing spinors invariant. We may also perform null rotations,
about $e^+$, of the $e^-$, $e^9$ and $e^8$. These do not fix the
individual Killing spinors, but they do preserve the four dimensional
subspace spanned by the Killing spinors. These null rotations lie in
an $\mbb^2$ subgroup of the isotropy group,
$(Spin(7)\ltimes\mbb^8)\times\mbb$, of $\e$. A general element of this
$\mbb^2$, in its spinorial representation, is
\be
1+p\G^{+8}+q\G^{+9}.
\een
The action of the full $(Spin(7)\ltimes\mbb^8)\times\mbb$ on the
spacetime basis is given in \cite{gaunt}.

We will also exploit the fact that the differential equations for the
spinorial components imply that all three additional Killing spinors must have
$v_8=\mbox{constant}$. Therefore, by taking linear combinations with
constant coefficients, we may always arrange that only one of the
additional Killing spinors has $v_8\neq0$. By assumption, at least
one of the remaining pair has $g\neq0$. Denote this spinor by $\eta$:
\be
\eta=(f^{\prime}+u_8^{\prime}\G^8+g^{\prime}\G^-)\e,
\een
where $g^{\prime}\neq0$ and without loss of generality we can always
choose $g^{\prime}>0$. Now we can make a specific choice of frame, by
acting on the Killing spinors with
\be
1+\frac{u_8^{\prime}}{g^{\prime}}\G^{+8}+\frac{f^{\prime}}{g^{\prime}}\G^{+9},
\een
so that in the new frame,
\be
\eta=g^{\prime}\G^-\e.
\een
With this choice of frame, by examining the differential equations for
the components of $\eta$, we find the additional conditions on the
spin connection:
\bea
\o_{-+9}=\o_{++9}=\o_{99+}=\o_{i9+}&=&\o_{++8}=\o_{9+8}=\o_{i8+}=0,\nn
\o_{8-+}&=&-\o_{-+8},
\eea
and the differential equations for $g^{\prime}$ are
\bea
\pa_-g^{\prime}=\pa_+g^{\prime}&=&\pa_9g^{\prime}=\pa_Ag^{\prime}=0,\nn
\pa_8\log g^{\prime}&=&-\o_{8-+}.
\eea
Since $\e$ and $\eta$ are stabilised by a common Spin(7) subgroup of
Spin(1,10), we have deduced that all spacetimes with maximal $G_2$
supersymmetry also admit a Spin(7) structure. The conditions on the
spin connection associated with the existence of a Spin(7) structure
were solved in \cite{spin7}; the most general local metric may always be put in
the form  
\bea
ds^2&=&(g^{\prime}(x))^2\Big(2[du+\lambda(x)_Mdx^M][dv+\v(x)_Ndx^N]+[dz+\s(x)_Mdx^M]^2\Big)\nn\label{110}&+&(g^{\prime}(x))^{-1}h_{MN}(x)dx^Mdx^N,
\eea
where $h_{MN}$ is a metric of Spin(7) holonomy and $d\lambda$, $d\v$
and $d\s$ are two-forms in the $\mathbf{21}$ of Spin(7). The spacetime
basis is given by 
 \bea
e^+&=&g^{\prime2}(du+\lambda),\nn e^-&=&dv+\v,\nn
e^9&=&g^{\prime}(dz+\s),\nn e^i&=&g^{\prime-1/2}\hat{e}^i(x)_Mdx^M,
\eea
where $\hat{e}^i$ are the achtbeins for $h$. Thus our task reduces to
imposing the additional constraints on the spin connection implied by
the existence of the $N=4$ $G_2$ structure on the metric (\ref{110}),
and then determining the remaining Killing spinors. The spin
connection for a metric of the form of (\ref{110}) was calculated in
\cite{gaunt}. It may be readily verified that having solved the
conditions for a Spin(7) structure, the only remaining conditions
on the spin connection for an $N=4$ $G_2$ that we have to solve are
\bea
\o_{ij-}=\o_{ij+}&=&\o_{ij9}=0,\nn
\o_{88A}&=&\o_{AB8}=0,\nn
\o_{8AB}&=&\o_{8AB}^{\mathbf{14}},
\eea
together with the differential equations for the spinorial
components. 
\\\\Firstly, the
conditions 
\be
\o_{ij-}=\o_{ij+}=\o_{ij9}=0
\een
imply that
\be
d\lambda=d\v=d\s=0,
\een
so by an $x$-dependent shift of the coordinates $u,v,z$ we may always
set $\lambda=\v=\s=0$ locally. To proceed, let us introduce
coordinates $w,y^{\bar{A}}$ such that the vector $\hat{e}^8$ is
\be
\hat{e}^8=\frac{\pa}{\pa w},
\end{equation}
and the achtbeins for $h$ are
\bea
\hat{e}^8&=&(dw+\rho(w,y)_{\bar{A}}dy^{\bar{A}}),\nn    
\hat{e}^A&=&\hat{e}^A_{\bar{A}}(w,y)dy^{\bar{A}}.
\eea
There are two distinct cases to consider: $\pa_8g^{\prime}=0$, and
$\pa_8g^{\prime}\neq0$. \\\\\paragraph{Case (i), $\pa_8g^{\prime}=0$.} If
$\pa_8g^{\prime}=0$, then without loss of generality we may take
$g^{\prime}=1$. We have $\o_{8-+}=0$, and $e^i=\hat{e}^i$. The
condition $\o_{88A}=0$ becomes
\be
(\pa_w\rho)_A=0,
\een
so $\rho=\rho(y)$. Then $\o_{AB8}$ is given by
\be
\o_{AB8}=\Psi_{(AB)}+\frac{1}{2}d\rho_{AB},
\een
where
\be
\Psi_{AB}=\d_{AC}(\pa_we^C)_B.
\een
Hence
\be
\Psi_{(AB)}=d\rho_{AB}=0,
\een
so locally we may set $\rho=0$. Next,
$\o_{8AB}=\o_{8AB}^{\mathbf{14}}$ implies that
\be
\Psi_{AB}=\Psi_{AB}^{\mathbf{14}},
\een
so that $\Psi$ is a two-form in the adjoint of $G_2$. This means that
the $w$-dependence of the $e^A$ is pure gauge. Under a $G_2$ rotation $Q$,
the $e^A$ transform according to
\be
e^A\rightarrow(e^A)^{\prime}=Q^A_{\;\;B}e^B.
\een
By performing a $w$-dependent $G_2$ rotation, we may choose the frame
so that
$\Psi_{AB}=0$, while leaving all four Killing spinors invariant. Thus
when $g^{\prime}=1$, the general local bosonic solution is the direct product of
$\mbb^{1,3}$ with a seven-manifold $\mathcal{M}$:
\be
ds^2=2dudv+dz^2+dw^2+h_{\bar{A}\bar{B}}(y)dy^{\bar{A}}dy^{\bar{B}}.
\een
Finally, requiring that the eight dimensional base space has Spin(7) holonomy implies that
\be
\o_{ABC}=\o_{ABC}^{\mathbf{14}},
\een
so $\mathcal{M}$ must have $G_2$ holonomy. The four-form
vanishes, and the Einstein equations are identically satisfied. Finally, we may
always choose the four Killing spinors to be
\be
\e,\;\;\G^8\e,\;\;\G^-\e,\;\;\G^{-8}\e.
\een

\paragraph{Case (ii): $\pa_8g^{\prime}\neq0$.} Now suppose that
$\pa_8g^{\prime}\neq0$. In this case, it is convenient to convert 
the outstanding conditions on the spin connection from the $e^i$ frame
to the conformally rescaled $\hat{e}^i$ frame. They become
\bea
\hat{\o}_{88A}&=&0,\nn
\hat{\o}_{8AB}&=&\hat{\o}_{8AB}^{\mathbf{14}},\nn
\hat{\o}_{AB8}&=&\frac{1}{2}\d_{AB}\pa_8\log g^{\prime},
\eea
and the condition that the conformally rescaled base has Spin(7)
holonomy is
\be
\hat{\o}_{ijk}^{\mathbf{7}}=0.
\een
As before, $\hat{\o}_{88A}=\hat{\o}_{[AB]8}=0$ implies that locally we
may set $\rho=0$. Hence, from the differential equations for
$g^{\prime}$, 
\be
g^{\prime}=g^{\prime}(w).
\een
Again as before,
$\hat{\o}_{8AB}=\hat{\o}_{8AB}^{\mathbf{14}}$ implies that
$\Psi_{[AB]}$ is pure gauge, and may be eliminated by a $G_2$
rotation. Then $\hat{\o}_{AB8}=\frac{1}{2}\d_{AB}\pa_8g^{\prime}$
implies that
\be
\Psi_{AB}=\frac{1}{2}\d_{AB}\pa_w\log g^{\prime}.
\een
This equation fixes the $w$-dependence of the $\hat{e}^A$ to be
\be
\hat{e}^A=g^{\prime1/2}(w)\tilde{e}^A(y),
\een
so the eight-metric in the $\hat{e}^8$, $\hat{e}^A$ frame becomes
\be
ds^2=dw^2+g^{\prime}(w)\tilde{h}_{\bar{A}\bar{B}}(y)dy^{\bar{A}}dy^{\bar{B}}.
\een
The final condition we have to impose is that this is a metric of
Spin(7) holonomy, $\hat{\o}_{ijk}^{\mathbf{7}}=0$. This condition is
equivalent to
\be
\d_{AB}\pa_w\log g^{\prime}=\hat{\Phi}_B^{\;\;CD}\hat{\o}_{ACD},
\een
where
\be
\hat{\Phi}=g^{\prime3/2}\Phi.
\een
By separating the $w$-dependence of both sides, there must exist some constant $R^{-1}$ such that
\be
\pa_wg^{\prime}=R^{-1}g^{\prime1/2},
\een
while in the conformally rescaled $\tilde{e}^A$ frame,
\be\label{111}
R^{-1}\d_{AB}=\tilde{\Phi}_B^{\;\;CD}\tilde{\o}_{ACD},
\een
and
\be
\tilde{\Phi}=\Phi=g^{\prime-3/2}\hat{\Phi}.
\een
So 
\be
g^{\prime}=\frac{1}{4}R^{-2}w^2,
\een
and (\ref{111}) is the statement that $\tilde{h}_{\bar{A}\bar{B}}$ is a metric
of weak $G_2$ holonomy on the seven-manifold $\mathcal{M}$ which is
spanned by the $\tilde{e}^A$; for more details on weak $G_2$
manifolds, see, for example, \cite{bilal}. This implies that the eleven dimensional
metric is locally the direct product of $AdS_4$ (with $AdS$ length
$R$) with a seven manifold of weak $G_2$ holonomy. Defining
$w=r^{-1/2}$, and performing a constant rescaling of the coordinates
$u$, $v$ and $z$, the eleven-metric takes the form
\be
ds^2=\frac{R^2}{r^2}(2dudv+dz^2+dr^2)+\tilde{h}_{\bar{A}\bar{B}}(y)dy^{\bar{A}}dy^{\bar{B}}.
\end{equation}
The four-form is given by
\be
F=-3R^{-1}e^+\wedge e^-\wedge e^9\wedge e^8.
\een
The four Killing spinors are  
\be
\e,\;\;(4R^2r)^{-1}\G^-,\;\;(-2vR^{-1}+zr^{-1}\G^-+\G^{-8})\e,\;\;(z+r\G^8+ur^{-1}\G^-)\e.
\end{equation}
The Bianchi identity is identically satisfied. This is just the
standard Freund-Rubin solution. Together with case (i),
this provides the general local bosonic solution of eleven dimensional
supergravity admitting four $G_2$ invariant Killing spinors.

\section{All spacetimes admitting a $(G_2\ltimes\mbb^7)\times\mbb^2$
  structure}
In this section, all spacetimes admitting a
$(G_2\ltimes\mbb^7)\times\mbb^2$ structure will be classified. As was
shown in \cite{nullstructure}, all $(G_2\ltimes\mbb^7)\times\mbb^2$
structures in eleven dimensions are defined by a pair of Killing
spinors; in addition to $\e$, the second Killing spinor may always be
taken to be
\be
(f+u\G^8)\e.
\een
We will now solve the equations of section two for this choice of
additional Killing spinor.

\subsection{The constraints for $(G_2\ltimes\mbb^7)\times\mbb^2$
  supersymmetry}
With the above choice of Killing spinor, equations (\ref{d-fA}),
(\ref{d-g}), (\ref{d9g}), (\ref{d-v}), (\ref{div}), (\ref{d-gA}) and
(\ref{d9gA}) are identically satisfied. Equations (\ref{dig}) and
(\ref{digA}) imply that
\bea
\ot_{i8-}&=&0,\nn
\ot_{iA-}-\frac{1}{3}\ot_{8j-}\ja&=&0,
\eea
and hence that
\be
\o_{ij-}=0,
\een
which from the $N=1$ conditions implies that
\be
F_{-98A}=F_{-9AB}=0.
\een
Then equation (\ref{d9v}) is satisfied, and the $A$ component of
(\ref{d-u}) implies that
\bea
\o_{-A8}&=&0,\nn
\o_{-AB}&=&\o_{-AB}^{\mathbf{14}}.
\eea\\\\Next, from the $A$ component of (\ref{d+u}), we find that
\be\label{ookoo}
F^{\mathbf{21}}_{+9A8}=4\os_{+A8}-2\ot_{+A8}.
\een
The flux component $F_{+9A8}$ is given by
$F_{+9A8}=F^{\mathbf{7}}_{+9A8}+F^{\mathbf{21}}_{+9A8}$. From the
$N=1$ condition on $F^{\mathbf{7}}_{+9ij}$, we find that
\be
F_{+9A8}=6\os_{+A8}-2\ot_{+A8}=-\o_{+ij}\phi^{ij}_{\;\;\;\;A8}=-\Phi_{A}^{\;\;\;BC}\o_{+BC}.
\een 
Under $G_2$, $F^{\mathbf{21}}_{+9AB}$ contains both a $\mathbf{7}$ and
a $\mathbf{14}$ part. The $\mathbf{7}$ part is determined by
(\ref{ookoo}). To see this, notice that
\be
F^{\mathbf{21}}_{+9A8}=\frac{1}{2}\Phi_A^{\;\;\;BC}F^{\mathbf{21}}_{+9BC}.
\een
Inverting this, we get
\be
(F^{\mathbf{21}}_{+9BC})^{\mathbf{7_{G2}}}=\frac{1}{3}F^{\mathbf{21}}_{+9A8}\Phi^A_{\;\;\;BC}=\frac{1}{3}(4\os_{+A8}-2\ot_{+A8})\Phi^A_{\;\;\;BC}, 
\een
where $(F^{\mathbf{21}}_{+9BC})^{\mathbf{7_{G2}}}$ denotes the
projection of $F^{\mathbf{21}}_{+9BC}$ on the $\mathbf{7}$ of
$G_2$. Similarly, for the flux components in the $\mathbf{7}$ of
Spin(7), we find that
\be
F^{\mathbf{7}}_{+9BC}=-F^{\mathbf{7}}_{+9A8}\Phi^A_{\;\;\;BC}=-2\os_{+A8}\Phi^A_{\;\;\;BC}.
\een
Hence, the part of $F_{+9BC}$ in the $\mathbf{7}$ of $G_2$ is given by
\be
F^{\mathbf{7_{G2}}}_{+9BC}=-\frac{2}{3}(\os_{+A8}+\ot_{+A8})\Phi^A_{\;\;\;BC}=-\frac{2}{3}\o_{+A8}\Phi^A_{\;\;\;BC}.
\een
The components $F^{\mathbf{14}}_{+9AB}$ in the $\mathbf{14}$ of $G_2$
drop out and are unconstrained. \\\\Next consider equation
(\ref{d+fA}). This gives
\be
\frac{16}{7}\o_{+9A}=F^{\mathbf{48}}_{+8ij}J^{Aij}=-F^{\mathbf{48}}_{+8BC}\Phi_A^{\;\;\;BC},
\een
or, equivalently,
\be
(F^{\mathbf{48}}_{+8BC})^{\mathbf{7_{G2}}}=-\frac{8}{21}\o_{+9A}\Phi^A_{\;\;\;BC}.
\een
Since from the $N=1$ conditions,
$F^{\mathbf{8}}_{+8BC}=-\frac{6}{21}\o_{+9A}\Phi^A_{\;\;\;BC}$, we
obtain
\be
F^{\mathbf{7_{G2}}}_{+8BC}=-\frac{2}{3}\o_{+9A}\Phi^A_{\;\;\;BC}.
\een
Under $G_2$, $F_{+ABC}=F^{\mathbf{8+48}}_{+ABC}$ contains $\mathbf{1}$,
  $\mathbf{7}$ and $\mathbf{27}$ parts. We may extract the different
pieces by contracting with $\Phi_{ABC}$ and $\Upsilon_{ABCD}$. Since 
\be
\Phi^{ABC}F^{\mathbf{48}}_{+ABC}=\phi^{ijk}_{\;\;\;\;\;8}F^{\mathbf{48}}_{+ijk}=0,
\een
the singlet is given by
\be
F^{\mathbf{1}}_{+ABC}=\frac{2}{7}\Phi_{ABC}\o_{+98}.
\een
Contracting $F^{\mathbf{8}}_{+ABC}$ with $\Upsilon$, we get
\be
F^{\mathbf{8}}_{+ABC}\Upsilon^{ABC}_{\;\;\;\;\;\;\;\;D}=\frac{48}{7}\o_{+9D}.
\een
We also find
\be
F^{\mathbf{48}}_{+ABC}\Upsilon^{ABC}_{\;\;\;\;\;\;\;\;D}=F^{\mathbf{48}}_{+ijk}\phi^{ijk}_{\;\;\;\;\;D}+3\Phi_D^{\;\;BC}F^{\mathbf{48}}_{+8BC}=-\frac{48}{7}\o_{+9D},
\een
so 
\be
F^{\mathbf{7}}_{+ABC}=0.
\een
The components $F^{\mathbf{27}}_{+ABC}$ drop out and are
unconstrained. \\\\Next, from the $8$ component of (\ref{d+v}), we
find
\be\label{uk}
\o_{+9-}=\frac{2}{3}\oy_{889}.
\een
Since from the $N=1$ conditions
$\o_{(ij)9}=\oy_{ij9}+\frac{1}{2}\d_{ij}\o_{+9-}$, this becomes
\be
\o_{889}=2\o_{+9-}=\o^A_{\;\;A9}.
\een
From the $A$ component of (\ref{d+v}),
\be\label{999}
2\os_{A89}-2\ot_{A89}-\oy_{A89}=0.
\een
From the $8$ component of (\ref{difA}), we obtain
\be\label{9999}
3\os_{A89}+\ot_{A89}-\frac{1}{2}\oy_{A89}=0.
\een 
The $B$ component is
\be\label{uj}
-\frac{3}{28}\d_{AB}\o_{+9-}+\frac{1}{4}F^{\mathbf{27_{Spin(7)}}}_{B8CD}\Phi^{ACD}+\os_{8C9}\Phi^C_{\;\;AB}+\ot_{AB9}-\frac{1}{2}\oy_{AB9}=0.
\een
The part antisymmetric on $A,B$, together with (\ref{999}) and
(\ref{9999}), implies that
\be
\o_{8A9}=\o_{A89}=\o_{[AB]9}=0.
\een
The $N=1$ conditions then imply that
\be
F_{+-A8}=F_{+-AB}=\os_{98A}=0.
\een
The trace on $A,B$ of (\ref{uj}) vanishes as a consequence of
(\ref{uk}). Defining 
\be
\o_{AB9}^{\mathbf{27}}=\o_{(AB)9}-\frac{1}{7}\d_{AB}\o^C_{\;\;C9},
\een
the traceless symmetric part of (\ref{uj}) becomes
\be
\frac{1}{2}F^{\mathbf{27_{Spin(7)}}}_{B8CD}\Phi^{ACD}-\o^{\mathbf{27}}_{AB9}=0.
\een
Hence
\be
F^{\mathbf{27_{Spin(7)}}}=-\frac{1}{4}\o^{\mathbf{27}}_{AB9}(J^A\wedge
J^B-\frac{1}{7}\d^{AB}J^C\wedge J^C).
\een
Then the $A$ component of (\ref{d9u}) implies that
\bea
\o_{98A}&=&0,\nn
\o_{9AB}&=&\o_{9AB}^{\mathbf{14}}.
\eea
It is now straightforward to verify that these conditions, together
with the $N=1$ constraints, imply that $F_{8ABC}$ and $F_{ABCD}$ are
determined as follows:
\bea
F^{\mathbf{1}}_{8ABC}&=&-\frac{6}{7}\Phi_{ABC}\o_{+9-},\nn
F^{\mathbf{7}}_{8ABC}&=&0,\nn
F^{\mathbf{27_{G2}}}_{8ABC}&=&-3\Phi_{[AB}^{\;\;\;\;\;\;D}\o_{C]D9}^{\mathbf{27}},\nn
F_{ABCD}&=&0.
\eea\\\\Next, from (\ref{d+g}), we find
\be
\o_{998}=2\o_{8-+}.
\een
Equations (\ref{d9fA}) and (\ref{d+gA}) imply that
\bea
\o_{99A}&=&2\o_{A-+},\nn
((\os_{[8AB]})^{\mathbf{48}})^{\mathbf{7_{G2}}}&=&\frac{2}{21}\Phi_{AB}^{\;\;\;\;\;C}\o_{C-+},
\eea
where $((\os_{[8AB]})^{\mathbf{48}})^{\mathbf{7_{G2}}}$ denotes the
projection of $(\os_{[8AB]})^{\mathbf{48}}$ on the $\mathbf{7}$ of
$G_2$. The $8A$ component of (\ref{diu}) reads
\be
\Phi_A^{\;\;\;BC}\o_{8BC}=0,
\een
so that
\be
\o_{8AB}=\o_{8AB}^{\mathbf{14}}.
\een
The $AB$ component of (\ref{diu}) is
\be\label{W2}
((\os_{[8AB]})^{\mathbf{48}})^{\mathbf{14_{G2}}}=-\frac{1}{12}\Phi_B^{\;\;\;CD}\o_{ACD},
\een
where $((\os_{[8AB]})^{\mathbf{48}})^{\mathbf{14_{G2}}}$ denotes the
projection of $(\os_{[8AB]})^{\mathbf{48}}$ on the $\mathbf{14}$ of
$G_2$.
Recall that for a $G_2$ structure in seven dimensions, the intrinsic
(con)torsion is specified by $\o^{\mathbf{7_{G2}}}_{ABC}$, and decomposes into four
modules,
\bea
T\in\Lambda^1\otimes
g_2^{\perp}&=&\mathcal{W}_1\oplus\mathcal{W}_2\oplus\mathcal{W}_3\oplus\mathcal{W}_4,\nn
\mathbf{7}\times\mathbf{7}&=&\mathbf{1}+\mathbf{14}+\mathbf{27}+\mathbf{7}.
\eea
Equation (\ref{W2}) states that $\o_{ABC}^{\mathbf{7_{G2}}}$, the
projection on $B,C$ of $\o_{ABC}$ onto the $\mathbf{7}$ of $G_2$, only
contains a $\mathcal{W}_2$ component.

At this point, we have solved all the purely algebraic equations
contained in the Killing spinor equation. However, we do not yet have
any explicit conditions on $\o_{88A}$, $\o_{AB8}$. These are in fact
contained in
\be\label{poiu}
\os_{[ijk]}=(\os_{[ijk]})^{\mathbf{8}}+(\os_{[ijk]})^{\mathbf{48}}.
\een
To extract them, we use the $N=1$ constraint on
$(\os_{[ijk]})^{\mathbf{8}}$, together with the conditions on
$(\os_{[8AB]})^{\mathbf{48}}$, $\o_{99i}$, $\o_{i-+}$ and $\o_{ABC}$
derived above. The $8AB$ component of (\ref{poiu}) reduces to
\be
\o_{[AB]8}=\Phi_{AB}^{\;\;\;\;\;C}(\o_{C-+}-\frac{1}{2}\o_{88C}).
\een
Observing that (\ref{W2}) implies that the totally antisymmetric part
of $\o_{ABC}^{\mathbf{7_{G2}}}$ vanishes,
$\o_{[ABC]}^{\mathbf{7_{G2}}}=0$, the $ABC$ component of (\ref{poiu})
becomes
\be
-\frac{1}{4}\Phi_{[AB}^{\;\;\;\;\;\;D}\o_{C]D8}=(\os_{[ABC]})^{\mathbf{48}}-\frac{1}{14}\Phi_{ABC}\o_{8-+}-\frac{1}{14}\Upsilon_{ABC}^{\;\;\;\;\;\;\;\;D}\o_{D-+}.
\een
The $\mathbf{1}$ part of this equation reads
\be
\o_{\;\;\;A8}^A=2\o_{8-+}.
\een
The $\mathbf{7}$ part gives
\be
\Phi_A^{\;\;\;BC}\o_{BC8}=0,
\een
and hence
\bea
\o_{[AB]8}&=&0,\nn
\o_{88C}&=&2\o_{C-+}.
\eea
The $\mathbf{27}$ part becomes 
\be
((\os_{[ABC]})^{\mathbf{48}})^{\mathbf{27_{G2}}}=-\frac{1}{4}\Phi_{[AB}^{\;\;\;\;\;\;D}\o_{C]D9}^{\mathbf{27}},
\een
where $((\os_{[ABC]})^{\mathbf{48}})^{\mathbf{27_{G2}}}$ denotes the
projection of $(\os_{[ABC]})^{\mathbf{48}}$ on the $\mathbf{27}$ of
$G_2$, and $\o_{AB8}^{\mathbf{27}}$ is defined in the same way as $\o_{AB9}^{\mathbf{27}}$. Now we may express the remaining flux components as follows:
\bea
F_{+-98}&=&F_{+-9A}=0,\nn
F^{\mathbf{7}}_{89AB}&=&2\Phi^{\;\;\;\;\;C}_{AB}\o_{C-+},\nn
F^{\mathbf{14}}_{89AB}&=&\Phi_A^{\;\;CD}\o_{BCD},\nn
F^{\mathbf{1}}_{9ABC}&=&\frac{6}{7}\Phi_{ABC}\o_{8-+},\nn
F^{\mathbf{7}}_{9ABC}&=&0,\nn
F^{\mathbf{27}}_{9ABC}&=&3\Phi_{[AB}^{\;\;\;\;\;\;D}\o_{C]D8}^{\mathbf{27}}.
\eea\\\\Finally, it is easily verified that the algebraic conditions
we have derived imply that the differential equations for the
spinorial components reduce to
\bea
\pa_{\m}f&=&0,\nn
\pa_{\m}u&=&0.
\eea
Hence given a solution of the conditions on the spin connection and
flux, we may always choose the second linearly independent Killing
spinor to be
\be
\G^8\e.
\een

\subsection{Summary}
Let us now summarise the above conditions for
$(G_2\ltimes\mbb^7)\times\mbb^2$ supersymmetry. In the spacetime basis
\be
ds^2=2e^+e^-+\d_{AB}e^Ae^B+(e^8)^2+(e^9)^2,
\een
the Killing spinors may be chosen to be
\be
\e,\;\;\G^8\e.
\een
The components of the spin connection are required to satisfy
\bea
\o_{(\m\v)-}=\o_{ij-}=\o_{i9-}&=&\o_{-9i}=\o_{-A8}=0,\nn
\o_{-AB}&=&\o_{-AB}^{\mathbf{14}},\nn
\o_{889}&=&2\o_{+9-}=\o_{\;\;A9}^A,\nn
\o_{8A9}=\o_{A89}&=&\o_{98A}=\o_{[AB]9}=0,\nn
\o_{9AB}&=&\o_{9AB}^{\mathbf{14}},\nn
\o_{998}&=&2\o_{8-+}=\o^A_{\;\;A8},\nn
\o_{99A}&=&2\o_{A-+}=\o_{88A},\nn
\o_{[AB]8}&=&0,\nn
\o_{8AB}&=&\o_{8AB}^{\mathbf{14}},\nn
\Phi_A^{\;\;\;CD}\os_{BCD}&=&(\Phi_{[A}^{\;\;\;CD}\os_{B]CD})^{\mathbf{14}},
\eea
where bold-face superscripts refer to $G_2$ representations and
$\os_{ABC}$ denotes the $\mathbf{7}$ projection of $\o_{ABC}$ on
$B,C$. Given a geometry satisfying the above conditions, the only
nonzero components of the flux are (again with bold face superscripts referring
to $G_2$ representations)
\bea
F_{+89A}&=&-\Phi_A^{\;\;\;BC}\o_{+BC},\nn
F^{\mathbf{7}}_{+9AB}&=&-\frac{2}{3}\o_{+C8}\Phi^C_{\;\;AB},\nn
F^{\mathbf{7}}_{+8AB}&=&\frac{2}{3}\o_{+C9}\Phi^C_{\;\;AB},\nn
F^{\mathbf{1}}_{+ABC}&=&\frac{2}{7}\Phi_{ABC}\o_{+98},\nn
F^{\mathbf{7}}_{89AB}&=&2\Phi^{\;\;\;\;\;C}_{AB}\o_{C-+},\nn
F^{\mathbf{14}}_{89AB}&=&\Phi_A^{\;\;CD}\o_{BCD},\nn
F^{\mathbf{1}}_{9ABC}&=&\frac{6}{7}\Phi_{ABC}\o_{8-+},\nn
F^{\mathbf{27}}_{9ABC}&=&3\Phi_{[AB}^{\;\;\;\;\;\;D}\o_{C]D8}^{\mathbf{27}},\nn
F^{\mathbf{1}}_{8ABC}&=&-\frac{6}{7}\Phi_{ABC}\o_{+9-},\nn
F^{\mathbf{27}}_{8ABC}&=&-3\Phi_{[AB}^{\;\;\;\;\;\;D}\o_{C]D9}^{\mathbf{27}},
\eea
together with $F^{\mathbf{14}}_{+9AB}$ and $F^{\mathbf{27}}_{+ABC}$
which drop out of the Killing spinor equations for $\e$ and $\G^8\e$
and are unconstrained by supersymmetry. Given a solution of these
conditions, it is sufficient to impose the Bianchi identity and
$E_{++}=Q_{+ij}=0$ to ensure that all field equations are satisfied.

\section{Conclusions}
In this work, a formalism for the exhaustive investigation of
all spacetimes with extended null supersymmetry in M-theory has been
provided. There is clearly much scope for the exploration of the
equations of section two. One way in which this could be
systematically pursued is by
working through all the possible structure groups of \cite{nullstructure}, case by
case. In particular, it should now be straightforward to classify,
reasonably explicitly, all spacetimes with maximal supersymmetry
consistent with a given structure group G, as was done for $G_2$
here. More challenging will be the classification of spacetimes with
less than maximal G-supersymmetry, particularly for spacetimes with
few supersymmetries and small structure groups, since in these cases
the Killing spinors will be quite generic and will involve many
functions, leading to a very complicated system of
equations. Spacetimes admitting generic identity structures will be
particularly difficult to cover in full generality. One would (perhaps
naively) expect spacetimes with more than sixteen supersymmetries to
be reasonably simple (indeed, it was shown in \cite{jose} that all
spacetimes with more than twenty-four are locally
homogeneous). However, such spacetimes necessarily admit an identity
structure, and it is precisely this case which seems to be hardest to treat
using G-structure methods. It would be interesting to get a
proper handle on these cases.

In addition to the general classification programme, it will be
interesting to use the formalism for highly targeted searches for
solutions of particular physical interest. One can insert, into the
equations of section two, any desired ansatz for the Killing spinors, the
metric, or both; used in this way, the refined
G-structure formalism of this paper is identical in spirit to a
completely general version of the algebraic Killing spinor techniques of
\cite{warner}. Of course, in any given context, the conditions for
supersymmetry may be recalculated using other techniques, most notably
algebraic Killing spinors; but the machinery of this paper covers {\it
  every} context where a null Killing spinor exists, and this
generality of the formalism ensures the generality of its
applications.   

The usefulness of the refined G-structure formalism stems from the
fact that it exploits the existence of a G-structure to extract and
package the
linearly independent first order PDEs contained within the Killing
spinor equation in a compact and (reasonably) tractable form. It
also provides a 
great deal of geometrical insight into the implications of
supersymmetry, through the relationship between the fluxes and the
intrinsic (con)torsion, for example. Nevertheless, one is still
 left with the task of solving these equations, and more
seriously, with solving the Bianchi identity and some subset of the field
equations. This will inevitably prove to be intractably difficult in
many cases. However, the general overview of the Killing spinor
equation provided by this approach should help reveal the directions
in which real progress can be made.

\section{Acknowledgements}
This work was supported by a Senior Rouse Ball Scholarship.

\appendix

\section{The integrability conditions}
Here we will analyse the integrability conditions for an arbitrary
additional Killing spinor $\eta$, assuming the existence of the
Killing spinor $\e$, and that the Bianchi identity for the four-form
is imposed on the solution of the Killing spinor equation. The
integrability condition we analyse is
\be\label{okoo}
[\G^{\v}[\md_{\m},\md_{\v}],f+u_i\G^i+\frac{1}{8}f^AJ^A_{ij}\G^{ij}+g\G^-+v_i\G^{-i}+\frac{1}{8}g^AJ^A_{ij}\G^{-ij}]\e=0,
\end{equation}   
where
\be
\G^{\v}[\md_{\m},\md_{\v}]=E_{\m\v}\G^{\v}+Q^{\v\s\t}\G_{\m\v\s\t}-6Q_{\m\v\s}\G^{\v\s}. 
\end{equation}
The algebraic conditions on the components of the field equations
implied by the existence of the Killing spinor $\e$ are given in
subsection \ref{oko}. To obtain the conditions implied by the
existence of an arbitrary additional Killing spinor, we impose the
projections satisfied by $\e$ to write each spacetime component of
(\ref{okoo}) as a manifest sum of basis spinors. The vanishing of each
coefficient then gives the conditions on the components of the field
equations. From the $-$ component, we find
\be\label{47}
v_iQ_{+-9}=0.
\een
From the $+$ component, we obtain
\be\label{48}
18Q_{+9i}u^i-gE_{++}=0,
\een
\be\label{49}
-18Q_{+ij}^{\mathbf{21}}u^j+E_{++}v_i=0,
\een
\be\label{50}
-36Q_{+9j}\ja u^i-9Q_{+ij}^{\mathbf{21}}f^BK^{BAij}+2E_{++}g^A=0,
\een
\be\label{51}
Q_{+9i}v^i=0,
\een
\be\label{52}
2Q_{+-9}u_i+2gQ_{+9i}-Q_{+ij}^{\mathbf{21}}v^j+2g^AQ_{+9j}\ja=0,
\een
\be\label{53}
4Q_{+9j}\ja v^i-Q_{+ij}^{\mathbf{21}}g^BK^{BAij}=0.
\een
The only independent condition obtained from the $9$ component is
\be\label{54}
Q_{+-9}u_i+gQ_{+9i}+Q_{+ij}^{\mathbf{21}}v^j+Q_{+9j}g^A\ja=0.
\een
Finally from the $i$ component, we get the new conditions
\be\label{55}
3gQ_{+ij}^{\mathbf{21}}+Q_{+9j}v_i+5Q_{+9i}v_j+\phi_{ij}^{\;\;\;kl}Q_{+9k}v_l+g^A\Big[2Q_{+jk}^{\mathbf{21}}J^{Ak}_{\;\;\;\;\;i}+Q_{+ik}^{\mathbf{21}}J^{Ak}_{\;\;\;\;\;j}\Big]=0,
\een
\be\label{56}
\Big[Q_{+-9}u_j+gQ_{+9j}+Q_{+jk}^{\mathbf{21}}v^k+Q_{+9k}g^AJ^{Ak}_{\;\;\;\;\;j}\Big]\ja+6v^jQ_{+k[i}^{\mathbf{21}}J^{Ak}_{\;\;\;\;\;j]}=0.
\een
Let us now analyse in detail the implications of these
equations. Combining (\ref{52}) and (\ref{54}), we find that
\be\label{57}
Q_{+ij}^{\mathbf{21}}v^j=0.
\een
Then combining (\ref{54}) and (\ref{56}), we find that
\be\label{58}
Q_{+kj}^{\mathbf{21}}\ja v^i=0.
\end{equation}
Since $v^j$, $\ja v^i$ are the components of eight linearly independent vectors,
(\ref{57}) and (\ref{58}) imply that 
\be
v_iQ_{+jk}^{\mathbf{21}}=0.
\end{equation}
Then by considering the cases $v_i=0$ and $v_i\neq0$ separately, from
(\ref{49}) we find that
\be
u^jQ_{+ij}^{\mathbf{21}}=v_iE_{++}=0.
\een
Similarly from (\ref{53}),
\be\label{59}
Q_{+9j}\ja v^i=g^AQ_{+ij}^{\mathbf{21}}K^{ABij}=0,
\een
and then (\ref{51}) yields
\be
v_iQ_{+9j}=0.
\een
Then given (\ref{47}) we see that if there exists an additional
Killing spinor with $v_iv^i\neq0$, it is sufficient to impose the
Bianchi identity, and all components of the field equations are
identically satisfied. \\\\Now suppose that $v_i=0$.
Since $Q_{+ij}$ only has components is the
$\mathbf{21}$, the coefficient of $g^A$ in equation (\ref{55}) only
has components in the $\mathbf{7}$ and $\mathbf{35}$ (and in fact, the
$\mathbf{7}$ part vanishes as a consequence of (\ref{59})). Hence
\be
gQ_{+ij}^{\mathbf{21}}=0.
\een
Then the vanishing of the $\mathbf{35}$ part of the coefficient of
$g^A$ in (\ref{55}), together with (\ref{59}), implies that
\be
g^AQ_{+ij}^{\mathbf{21}}=0.
\een
We have now solved all the integrability conditions with $v_i=0$, save
for equations (\ref{48}), (\ref{50}) and (\ref{52}). Consider first
the case $g^2+g^Ag^A\neq0$. Then if we {\it impose} $Q_{+-9}=0$,
equation (\ref{52}) becomes
\be
gQ_{+9i}+g^AQ_{+9j}\ja=0.
\een
Contracting with $Q_{+9}^{\;\;\;\;\;i}$ we find that $gQ_{+9i}=0$, and
contracting with $J^{Bij}Q_{+9j}$ we find that $g^AQ_{+9i}=0$. Hence
\be
(g^2+g^Ag^A)Q_{+9i}=0.
\een
Since $g^AQ_{+ij}^{\mathbf{21}}=0$, equations (\ref{48}) and
(\ref{50}) reduce to
\be
(g^2+g^Ag^A)E_{++}=0.
\een
So when there exists an additional Killing spinor with $v_i=0$,
$g^2+g^Ag^A\neq0$, it is sufficient to impose the Bianchi identity and
$Q_{+-9}=0$ to ensure that all field equations are
satisfied.\\\\Finally consider the case $v_i=g=g^A=0$. If
there exists an additional Killing spinor with $u_iu^i\neq0$, then from (\ref{48}), (\ref{50}) and (\ref{52}) it is
sufficient to impose $E_{++}=Q_{+ij}^{\mathbf{21}}=0$ in addition to the Bianchi
identity. If there exists an additional Killing spinor with
$v_i=g=g^A=u_i=0$, then it is sufficient to impose
$E_{++}=Q_{+-9}=Q_{+9i}=Q_{+ij}^{\mathbf{21}}=0$.

\end{document}